\documentclass[%
 aip,
rsi,%
 amsmath,amssymb,
 reprint,%
]{revtex4-1}
\usepackage[english]{babel}
\usepackage[utf8x]{inputenc}
\usepackage[T1]{fontenc}
\usepackage{subcaption}
\usepackage{soul}

\usepackage{amsmath}
\usepackage{graphicx}
\usepackage{epstopdf}
\usepackage[colorinlistoftodos]{todonotes}
\usepackage[colorlinks=true, allcolors=blue]{hyperref}

\begin{document}
\title{Experiments with arbitrary networks in time-multiplexed delay systems}

\author{Joseph D. Hart}
\affiliation{ 
Institute for Research in Electronics and Applied Physics, University of Maryland, College Park, Maryland 20742, USA
}%
\affiliation{ 
Department of Physics, University of Maryland, College Park, Maryland 20742, USA
}%
\author{Don C. Schmadel}
\affiliation{ 
Institute for Research in Electronics and Applied Physics, University of Maryland, College Park, Maryland 20742, USA
}%
\author{Thomas E. Murphy}
\affiliation{ 
Institute for Research in Electronics and Applied Physics, University of Maryland, College Park, Maryland 20742, USA
}%
\affiliation{Department of Electrical and Computer Engineering, University of Maryland, College Park, Maryland 20742, USA}
\author{Rajarshi Roy}
\affiliation{ 
Institute for Research in Electronics and Applied Physics, University of Maryland, College Park, Maryland 20742, USA
}%
\affiliation{ 
Department of Physics, University of Maryland, College Park, Maryland 20742, USA
}%
\affiliation{ 
Institute for Physical Science and Technology, University of Maryland, College Park, Maryland 20742, USA
}%

\date{\today}

\begin{abstract}
We report a new experimental approach using an optoelectronic feedback loop to investigate the dynamics of oscillators coupled on large complex networks with arbitrary topology. Our implementation is based on a single optoelectronic feedback loop with time delays. We use the space-time interpretation of systems with time delay to create large networks of coupled maps. Others have performed similar experiments using high-pass filters to implement the coupling; this restricts the network topology to the coupling of only a few nearest neighbors. In our experiment, the time delays and coupling are implemented on a field-programmable gate array, allowing the creation of networks with arbitrary coupling topology. This system has many advantages: the network nodes are truly identical, the network is easily reconfigurable, and the network dynamics occur at high speeds. We use this system to study cluster synchronization and chimera states in both small and large networks of different topologies.

\end{abstract}

\maketitle

\textbf{
Complex networks of coupled oscillators have proven to be systems that can display incredibly rich dynamical behaviors. Despite great theoretical advances in our understanding of coupled oscillator networks, it has proven difficult to design experiments that permit the study of dynamics on large networks with arbitrary topology. Here we present a new experimental approach that allows for the investigation of large networks of truly identical nodes with arbitrary topology. Our approach relies upon the space-time interpretation of systems with time delay in order to construct a network of coupled maps using a single nonlinear, time-delayed feedback loop. Our experiment provides a significant improvement over previous network experiments in terms of speed, ease of implementation, and cost. This paper describes our experimental approach and its capabilities, including the ability to study cluster synchronization and chimera states in both small and large networks.}

\section{Introduction}

Systems of coupled oscillators have been of great theoretical and practical interest since Huygens's observation of the anti-phase synchronization of two pendulum clocks \cite{huygens1897huygens}. Over the last 30 years, our understanding of coupled oscillators has grown from the two coupled pendula of Huygens to the synchronization \cite{pikovsky2003synchronization,pecora2015synchronization} and control \cite{sorrentino2007controllability} of nonlinear oscillators coupled together in complex networks of interactions. Recently, there has been an explosion of interest in patterns of network synchronization, such as chimera states \cite{abrams2004chimera,panaggio2015chimera} and cluster synchronization \cite{zhou2006hierarchical,ji2013cluster}(see, e.g., the Chaos special issue on Patterns of Network Synchronization \cite{abrams2016introduction}). 

However, experimental progress on complex networks has been slow compared to theoretical advances. For example, it took a decade from the initial prediction of chimera states in 2002 \cite{kuramoto2002coexistence} to their first experimental observations in 2012 \cite{hagerstrom2012experimental,tinsley2012chimera}. Only recently has there been significant progress in the experimental study of large networks. 

Many of the experiments on large networks of oscillators are limited in the types of coupling they can achieve. Electrochemical \cite{schmidt2014coexistence} and passively mode locked laser \cite{viktorov2014coherence} systems with nonlinear global coupling have been used to study chimera states. Electronic and optoelectronic systems with time delay have used the virtual space-time \cite{arecchi1992two} interpretation to implement large networks of coupled oscillators for the study of both chimera states \cite{larger2013virtual,larger2015laser} and reservoir computing \cite{appeltant2011information,paquot2012optoelectronic,larger2012photonic,antonik2017brain}; however, these systems so far have been restricted to networks with cyclic symmetry. Coupled metronomes have been used to study chimera states in hierarchical networks \cite{martens2013chimera,blaha2016symmetry}, but this approach is impractical for the implementation of networks with complex topologies. The authors are aware of only a few experiments that can be used to study large networks with arbitrary coupling topology. Chemical oscillators \cite{tinsley2012chimera,totz2015phase,wickramasinghe2013spatially,wickramasinghe2014spatially} and spatial light modulator feedback loops \cite{hagerstrom2012experimental,pecora2014cluster} have been used to study hundreds of coupled oscillators, but these experiments run on slow timescales, can be expensive and have inherent heterogeneities between nodes.

In this paper, we describe a new approach that allows for the experimental investigation of the dynamics of arbitrary networks of coupled maps. Our approach is similar to that of previous opto-electronic experiments \cite{kouomou2005chaotic} that use the space-time interpretation to create coupled node networks with a single time-delayed system \cite{appeltant2011information,larger2012photonic,larger2013virtual,larger2015laser,antonik2017brain}. However, these experiments are restricted to cyclically symmetric networks due to the time-invariant (in these cases, highpass) filtering in their electronics. We use optics and electronics designed to have no highpass filtering; this removes the few-nearest neighbor coupling that was always present in previous work. Additionally, we use a field-programmable gate array (FPGA) to perform time multiplexing in order to implement arbitrary coupling topologies between the nodes. This approach has the additional advantage that it makes it possible to obtain a network of truly identical nodes because all nodes utilize the same electronic and optical components. This eliminates the (impossible) requirement to build many individual identical systems, and replaces it with the much easier task of obtaining temporal stability. Our experimental system is low cost, high speed, and easily reconfigurable compared with other experiments on large networks with arbitrary topology.

Our system provides a versatile experimental platform to study dynamics on all sorts of networks. The coupling between nodes can be instantaneous and/or time-delayed. The network can be static, time-varying, or adaptive. The nodes are truly identical since they all use the same experimental apparatus; however, they can be made inhomogeneous by controlling node parameters in the FPGA for each node. The ability to control the inhomogeneity of the nodes should enable the experimental study of multi-layer networks \cite{boccaletti2014structure} and Asymmetry-Induced Synchronization \cite{nishikawa2016symmetric}. One can introduce targeted perturbations to individual nodes, which may be useful for testing basin stability \cite{menck2013basin,menck2014dead} and controllability \cite{liu2011controllability}.

The paper is organized as follows. In Section \ref{sec:singlenode}, we describe the experiment for a single nonlinear map and investigate its dynamics. In Section \ref{sec:twonodes}, we discuss how we create two coupled nonlinear maps using the space-time interpretation of time-delayed systems. In Section \ref{sec:networks}, we show how this same system can be used to implement larger networks of coupled maps, and we use it to study cluster and chimera states in both small and large networks of different topologies. 


\section{A single nonlinear map \label{sec:singlenode}}
\begin{figure}
\centering
\includegraphics[width=0.45\textwidth]{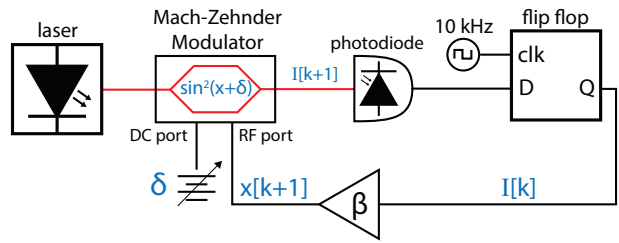}
\caption{\label{fig:singlenode}Schematic for the single nonlinear map. In our experiment, the flip flop and clock are implemented on a FPGA.}
\end{figure}

We first discuss the simplest version of the experiment: operation as a single nonlinear map. Figure \ref{fig:singlenode} provides a schematic of the experiment. Light from an 850 nm continuous-wave, fiber-coupled distributed feedback laser passes through an integrated LiNO\textsubscript{3} electro-optic intensity modulator ($V_{\pi,RF}$ = 1.85 V) and is converted into an electrical signal by a photoreceiver. A flipflop performs a sample-and-hold operation on the electrical signal at a rate of $F_s$ = 10 kHz. This is implemented on an FPGA (Altera Cyclone V GT), and the delay caused by the clocked sample-and-hold is much longer than the optical delay. This makes time discrete and decouples consecutive time steps. The electronic output of the flipflop is amplified then applied to the RF port of the electro-optic modulator. The modulator provides the nonlinearity. An independent power supply controls the DC bias of the electro-optic modulator.

This experiment can be described by the following nonlinear map:
\begin{equation}
x[k+1] = \beta I(x[k]),
\end{equation}
where $x = \pi v/2V_{\pi,RF}$ is the normalized voltage applied to the modulator at discrete time $k$, and $\beta$ is the normalized round-trip gain.
The normalized intensity of the light passing through the electro-optic modulator can be modeled as $I(x) = \sin^2(x+\delta)$, where $\delta \equiv \pi V_{dc}/2V_{\pi,dc}$ is the DC bias point of the electro-optic modulator. The sine-squared nonlinearity is intrinsic to all wave-interference devices, including our intensity modulator. With this experimental system, one could obtain any desired nonlinearity $I(x)$ by performing a nonlinear operation on $x$ in the FPGA. Alternatively, one could introduce a different nonlinearity by using different, sufficiently fast nonlinear optical or electronic components.

Figure 2a shows an experimentally measured bifurcation diagram of the single nonlinear map with $\delta=\pi/4$. Our experimental map can exhibit fixed point, periodic, and chaotic behaviors. A bifurcation diagram numerically simulated from Eq. (1) and shown in Fig. 2b, agrees well with the experiment and suggests that Eq. (1) is an accurate model.

If $\beta$ is too large, the voltage $v$ applied to the modulator may exceed the allowed limit for the device (for our modulator, the maximum voltage range is about 8V). This would limit the permissible values of $\beta$ to about 2.5$\pi$. However, since $\sin^2(\cdot)$ is $\pi$ periodic, we resolve this problem by calculating $x$ modulo $2\pi$ in the FPGA, allowing our experiment to operate at very high values of $\beta$.

\begin{figure}
\centering
\includegraphics[width=0.45\textwidth]{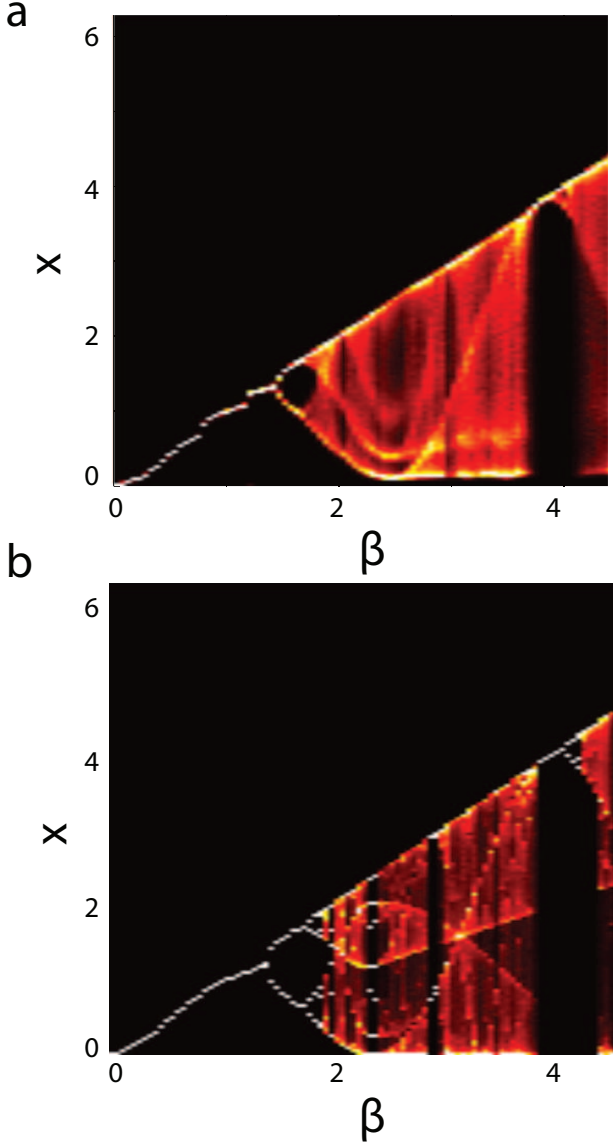}
\caption{\label{fig:bif_diagram} \textit{a}) Experimental bifurcation diagram of the single nonlinear oscillator depicted in Fig. 1. The DC bias $\delta$ was set to $\pi/4$. \textit{b}) Numerically simulated bifurcation diagram of the single nonlinear map described by Eq. 1. $\delta$ was set to $\pi/4$. 15000 samples were used for each value of $\beta$.}
\end{figure}

\section{Two coupled nonlinear maps \label{sec:twonodes}}
In order to explore the dynamics of coupled nonlinear systems, one can envision coupling two nominally identical nonlinear maps. Such a system can be modeled as
\begin{equation}
\label{eq:twonodes}
\begin{split}
x_0[n+1] &= \beta I(x_0[n]) + \sigma I(x_1[n])\\
x_1[n+1] &= \beta I(x_1[n]) + \sigma I(x_0[n]),
\end{split}
\end{equation}
where $\sigma$ is the coupling strength.

However, there are significant experimental difficulties in directly coupling two nonlinear maps. Two separate systems must be built and coupled; even then, the individual systems are only nominally identical. 

\begin{figure}
\centering
\includegraphics[width=0.45\textwidth]{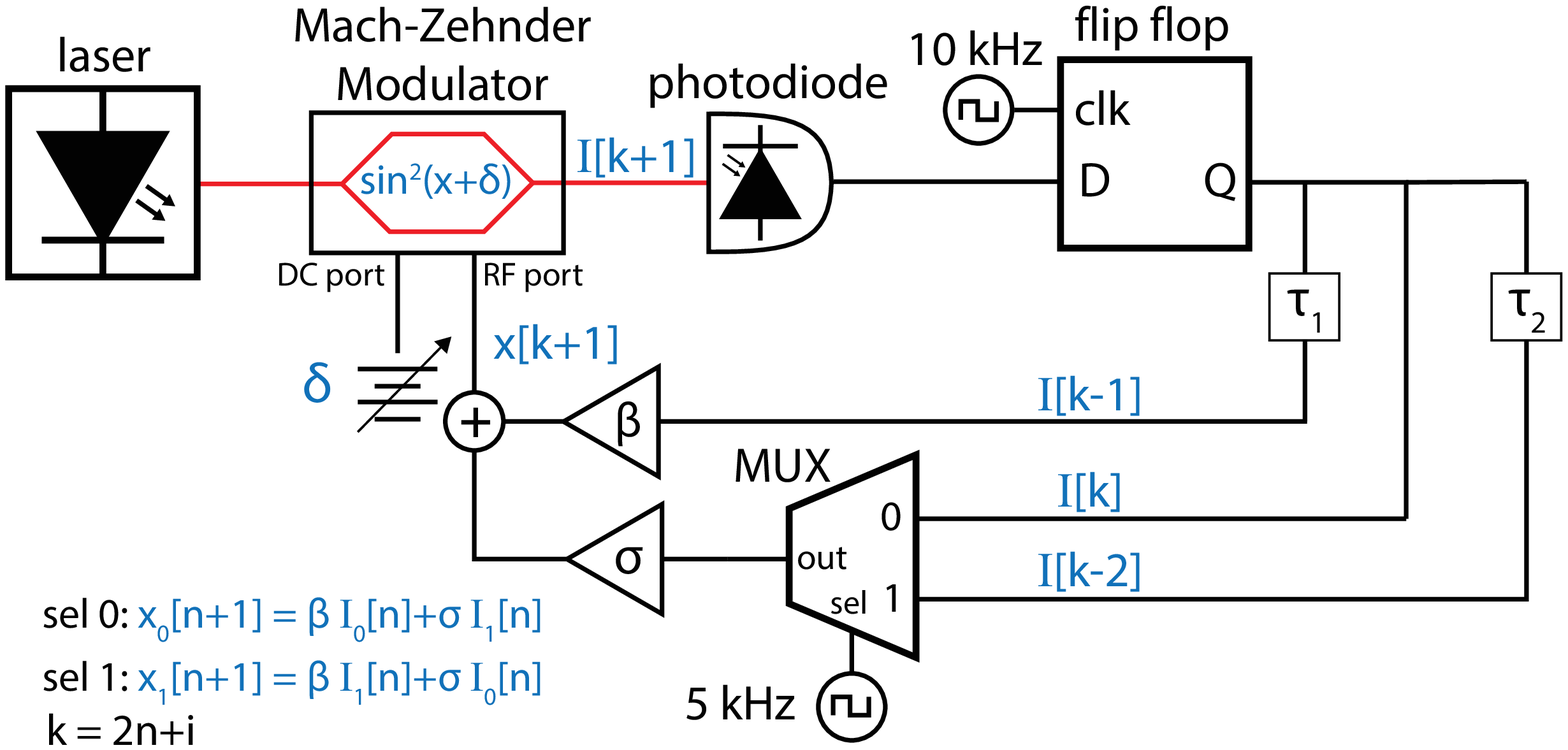}
\caption{\label{fig:twonodes} Illustration of a time-delayed implementation of two coupled oscillators. The standard time $k$ updates at a rate of 10 kHz, while the network time $n$ updates at a rate of 5 kHz. MUX denotes a multiplexer that outputs $I[k]$ on when $k$ is even and $I[k-2]$ when $k$ is odd. $\tau_1$ = 1 time step (100 $\mu$s) and $\tau_2$ = 2 time steps (200 $\mu$s).}
\end{figure}

One can, however, modify the single map described above to realize two coupled maps by adding time-delays as shown in Fig. \ref{fig:twonodes}. This is our approach. In this setup, the single oscillator with time-delays and a multiplexer (MUX) functions as two individual nonlinear maps coupled together. Every even time step $k$ of the full system is interpreted as an iteration of node 0, and every odd time step $k$ of the full system is interpreted as an iteration of node 1. This system can also be modeled by Eq. \ref{eq:twonodes}, but requires the assembly of only one experimental apparatus. Further, each of the two nodes are identical, since they utilize the same electronic and optical components. The individual maps now update at half the rate since the measurement of $I$ is performed serially.

In practice, the flip flop, time delays, time multiplexing, multiplication by $\beta$ and $\sigma$, and addition of the self-feedback and coupling components are all implemented on an FPGA. This allows for a compact and flexible implementation of the experiment: The experiment can be switched between the single nonlinear map described in section \ref{sec:singlenode} and the coupled map system simply by reconfiguring the FPGA.

\begin{figure}
\centering
\includegraphics[width=0.45\textwidth]{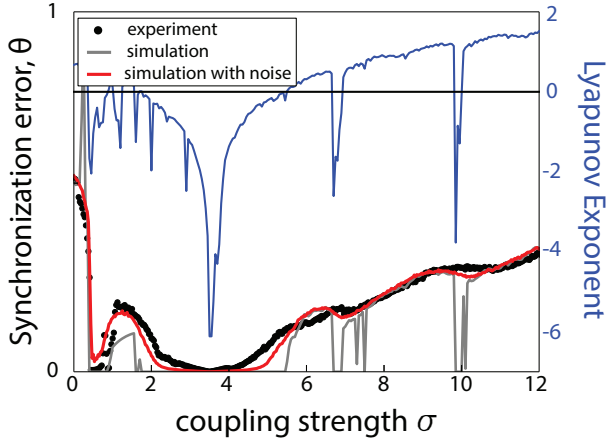}
\caption{\label{fig:two_node_sync}Synchronization of two nodes as a function of $\sigma$ for $\beta=3.5$ and $\delta=\pi/4$. The coupling is symmetric and bidirectional, as in Eq. \ref{eq:twonodes}. The gray line indicates the synchronization error in a noiseless simulation, while the red line indicates the synchronization error in a simulation in which Gaussian noise with standard deviation 0.02 was added to the intensities $I$ at each time step. The blue line indicates the numerically computed Lyapunov exponent corresponding to perturbations transverse to the synchronization manifold.}
\end{figure}

Figure \ref{fig:two_node_sync} shows the measured and simulated root-mean-square (RMS) synchronization error of the coupled map system depicted in Fig. \ref{fig:twonodes} and described by Eq. \ref{eq:twonodes}. We define the RMS synchronization error as 

\begin{equation}
\label{eq:syncerror}
\theta \equiv \Big(\frac{\langle(x_0(t)-x_1(t))^2\rangle}{\langle x_0(t)^2+x_1(t)^2\rangle}\Big)^{1/2},
\end{equation}

where $\langle\cdot\rangle$ denotes an average over time. $\theta$ is zero in the case of a completely synchronized solution and approaches 1 in the limit that the $x_0$ and $x_1$ are uncorrelated.

One can determine the stability of the synchronized state by linearizing about the synchronized solution $ x_s[n]\equiv x_0[n]=x_1[n]$ and calculating the Lyapunov exponent of the variational equation for the perturbations transverse to the synchronization manifold. The variational equation is
\begin{equation}
\label{eq:variational2node}
\Delta x_\perp[n+1] = (\beta-\sigma)\sin\big(2(x_s[n]+\delta)\big)\Delta x_\perp[n],
\end{equation}
where $\Delta x_\perp$ is a perturbation transverse to the synchronization manifold. We have calculated the Lyapunov exponent of Eq. \ref{eq:variational2node} as a function of $\sigma$ for fixed $\beta=3.5$; the results are shown in Fig. \ref{fig:two_node_sync}. The values of $\sigma$ with negative Lyapunov exponent correspond exactly to the values where we observe synchronization in the simulations without noise. Further, all of the $\sigma$ values where we observe synchronization in the experiment correspond to $\sigma$ with negative Lyapunov exponent; however there are some narrow regions of $\sigma$ with negative Lyapunov exponent where we do not observe synchronization in the experiment. This is due to noise in the experiment. Noise in the experiment comes from a variety of sources; some of these noise sources include discretization noise in the ADC and DAC, electronic noise in the DAC amplifier, and Johnson noise in the photoreceiver. Because the voltage applied to the modulator can have a DC component, significant electrical power can be dissipated in the modulator's 50$\Omega$ input resistor. This can result in heating, and so there are also fluctuations in the intensity at the output of the modulator. 

We model all of these sources of noise by applying additive Gaussian noise with standard deviation 0.02 to each normalized intensity $I_i$ at each time step:

\begin{equation}
\begin{split}
x_0[n+1]=\beta\big(I(x_0[n])+aR_0[n]\big)+\sigma \big(I(x_1[n])+aR_1[n]\big) \\
x_1[n+1]=\beta\big(I(x_1[n])+aR_1[n]\big)+\sigma \big(I(x_0[n])+aR_0[n]\big)
\end{split}
\end{equation}
where $R_i[n]$ are independent, identically distributed random variables taken from a Gaussian distribution with zero mean and unit variance.
The synchronization error from the simulation with $a=0.02$, shown in red in Fig. \ref{fig:two_node_sync}, shows good agreement with the experimentally measured result.

\section{\label{sec:networks}Complex networks of nonlinear maps}
\begin{figure}
\centering
\includegraphics[width=0.45\textwidth]{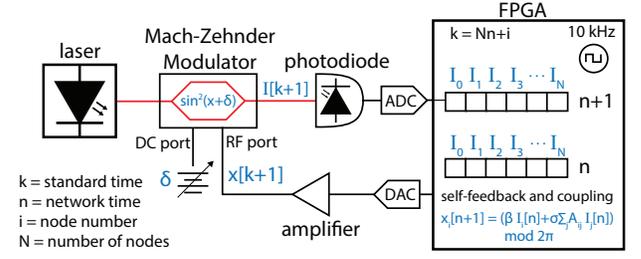}
\caption{\label{fig:FPGA}Schematic of the experimental apparatus for exploring complex networks of coupled nonlinear maps. The coupling is implemented in hardware on the FPGA. Data are taken by streaming the measured values of $I$ and the updated values of $x$ directly from the FPGA to a computer.}
\end{figure}

The experiment described in Sec. III can be adapted as shown in Fig. \ref{fig:FPGA} to create an arbitrary network of coupled nonlinear maps. The only modification is in the FPGA configuration. The normalized voltage output $x$ of the FPGA is now a linear combination of the previous normalized intensity $I$ measurements. This can be interpreted as a network of coupled oscillators according to the space-time interpretation described in ref. \onlinecite{arecchi1992two} and utilized in ref. \onlinecite{appeltant2011information}.

In the space-time interpretation, there are two ``time steps'': the standard time $k$ and the network time $n$. In particular, for a network of $N$ nodes, at standard time $k+1=N\cdot (n+1)+i$, the node $i$ is updated as follows: $x_i[n+1] = \beta I[N\cdot n+i] + \sigma\sum_jA_{ij}I[N\cdot n+j]$, where $n$ is the network time. $x_i[n]$ is the normalized voltage applied to the modulator at standard time $k=N\cdot n+i$. The FPGA then measures and records $I[N\cdot (n+1)+i]$, the normalized intensity that passes through the modulator. $I[N\cdot (n+1)+i]$ is stored in a shift register on the FPGA, and is used to update nodes in the future. The network updates at a rate of $F_s/N$.

The system can be modeled in network time as

\begin{gather}
\label{eq:network}
x_i[n+1]=\beta I(x_i[n]) + \sigma\sum_jA_{ij}I(x_j[n]) \\
I(x_i[n]) = \sin^2(x_i[n] + \delta).
\end{gather}
Eq. \ref{eq:network} clearly shows that this system can be understood to be a network of coupled maps.

We note that $A_{ij}$ can be a Laplacian matrix, which indicates diffusive coupling and can lead to negative values of $x_i$. This is not a problem because $x$ is determined from the intensities using signed arithmetic in the FPGA. We take advantage of the $\pi$-periodicity of the modulator to allow us to output only positive voltages $v$ from our unipolar digital-to-analog converter.

Similar optoelectronic experiments have used the space-time interpretation to realize networks for reservoir computing \cite{appeltant2011information,larger2012photonic,antonik2017brain} and to observe chimeras \cite{larger2015laser}. These systems use a time-invariant high-pass filter, which creates few nearest-neighbor coupling. In contrast, we are able to create arbitrary networks of coupled maps in a table-top experimental network by removing the high-pass filter and using a flipflop to sample-and-hold in the FPGA to obtain discrete time. Our nodes are maps; discrete time is a natural consequence of the space-time interpretation of systems with time delays \cite{arecchi1992two}. Further, since $\beta$ is controlled by the FPGA in our experiment, we are able to explore networks with either identical (by keeping $\beta$ constant in time) or non-identical nodes (by varying $\beta$ periodically in time).

\subsection{Chimeras in large ring network}
\begin{figure*}
\centering
\includegraphics[width=\textwidth]{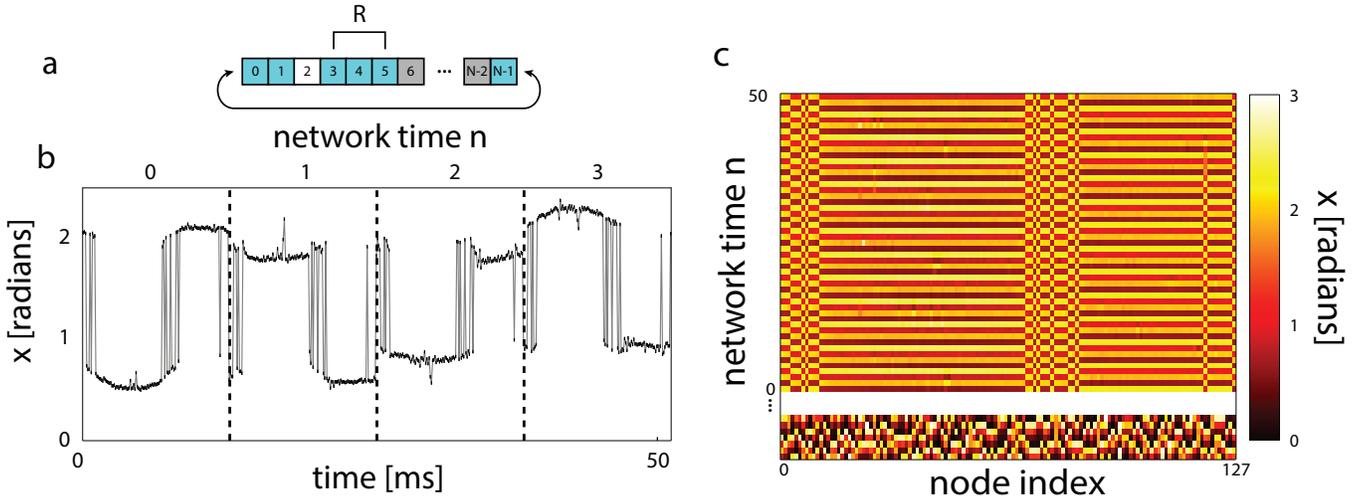}
\caption{\label{fig:ringchimera}\textit{a}) Depiction of the ring network topology of the system in which we observe a chimera state. $N$=128 nodes and $R$=52 nodes. Node 2 (white) is coupled to its $R$ nearest neighbors (blue) on either side. \textit{b}) Experimental time series of the chimera state. The state of each node is represented by a dot; the states of adjacent nodes are connected with a gray line to guide the eye. The vertical black dotted lines separate the network time steps. \textit{c}) Space-time network interpretation of \textit{b}. The random initial conditions before the coupling $\sigma$ is turned on is shown before network time $n$=0. Once $\sigma$ is turned on after a long transient (not shown), the system reaches the chimera state, which begins at $n$=0.}
\end{figure*}

In order to demonstrate that our experimental system can be used to study large networks, we now investigate chimera states in a ring network of  128 nodes. As shown in Fig. \ref{fig:ringchimera}a, each node is coupled to the $R$ nodes on either side of it with equal strength $\sigma$; this is the same coupling scheme used in ref. \onlinecite{hagerstrom2012experimental} to observe chimeras. The nodes are coupled in a Laplacian fashion. The network can be modeled by Eq. \ref{eq:network} with the adjacency matrix given by

  \[
    A_{ij}=\left\{
                \begin{array}{ll}
                  1\qquad\qquad\qquad 0<(j-i)\ \rm{modulo}\ N \leq R\\
      
                 0\qquad\qquad\qquad\ \ \ (j-i)\ \rm{modulo}\ N > R\\
                 -\sum_{j\neq i}A_{ij} \qquad\qquad\qquad i=j
                  
                \end{array}
              \right.
  \]

For this experiment, we set $\beta=$2.67, $\delta=$0, $N$=128, and $R$=52, which is a scaled version of the parameters used in ref. \onlinecite{hagerstrom2012experimental} At these values of $\beta$ and $\delta$, a single uncoupled map behaves chaotically. However, when coupled together with $\sigma=$0.0107, the chimera state displayed in Fig. \ref{fig:ringchimera}b is observed. In this state, all nodes have period-4 dynamics, but there are two regions of spatial coherence separated by two regions of spatial incoherence. The space-time network interpretation is shown in Fig. \ref{fig:ringchimera}c. The system starts with the oscillators uncoupled. They oscillate chaotically and independently, as shown. There is a long transient after the coupling is turned on (not shown). At the end of the transient, the system settles into the chimera state (shown starting at time $n=0$).

At a size of 128 nodes, the network updates at 10 kHz/128 $\approx$ 78 Hz, which is an order of magnitude faster than previous coupled map experiments implemented with a SLM\cite{hagerstrom2012experimental}. Another advantage of our experiment, as noted earlier, is that all the nodes are identical. This removes any question of potential inhomogeneities among oscillators leading to the chimera states.

Since this network exhibits cyclic symmetry, it might be possible to implement in the previous versions of experiments using time delays to implement virtual nodes \cite{appeltant2011information,larger2012photonic,larger2013virtual,larger2015laser,antonik2017brain} as discussed above; however, this has never been done. The chimeras that have been observed in these systems are qualitatively different than those observed here.

\subsection{Cluster synchronization in a 5 node network with symmetries}
\begin{figure*}
\centering
\includegraphics[width=\textwidth]{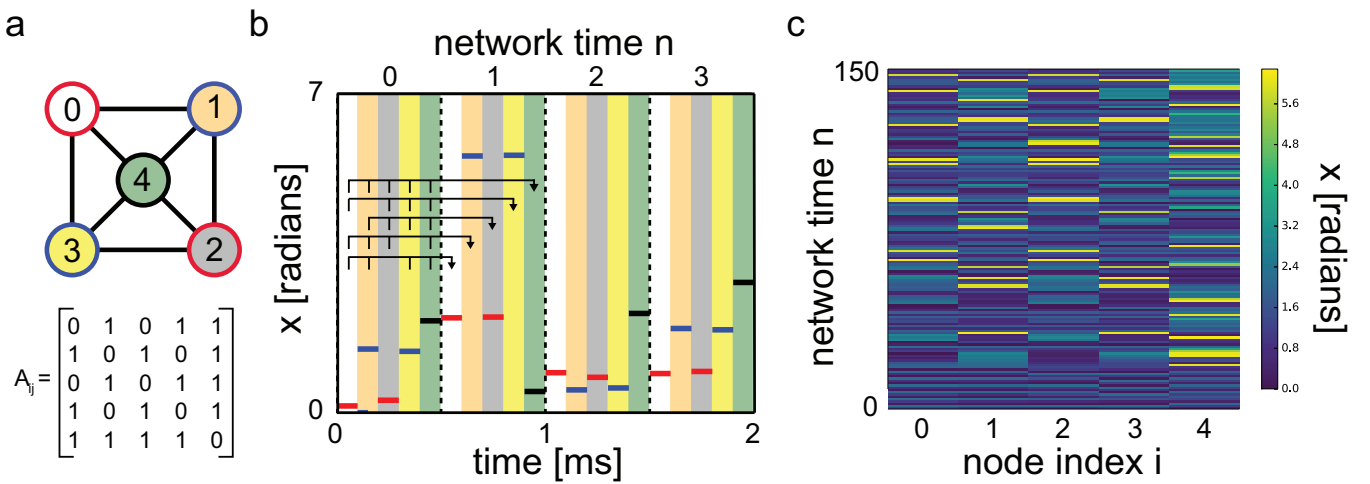}
\caption{\label{fig:5node}\textit{a}) 5 node network visualization and adjacency matrix. The color of the outline on each node indicates which synchronous cluster it belongs to. \textit{b}) Time series of the 5 node network in the cluster state depicted in \textit{a}. The background color of each time step indicates the node from \textit{a} that is represented by that time step. The coloring of the horizontal bar indicates the synchronous cluster to which that node belongs. The vertical black dotted lines indicate a full network time step $n$. The black arrows depict how the network structure is implemented by time multiplexing. \textit{c}) Space-time network interpretation of \textit{b}. }
\end{figure*}

In the previous section, we demonstrated that our experimental system can be used to create realizations of networks with large numbers of nodes. In order to illustrate that our approach can be used to realize networks with arbitrary topology, we implement the network shown in Fig. \ref{fig:5node}a. Previous experiments using time delays and time-invariant filters to implement cyclically symmetric networks \cite{appeltant2011information,larger2012photonic,larger2013virtual,larger2015laser,antonik2017brain} cannot be used to study this network. We reproduce a result from ref. \onlinecite{sorrentino2016complete}; this work explored the stability of cluster synchronization in a 5 node network with Laplacian coupling implemented on a SLM feedback system. We show that our time-delayed feedback network can also display cluster synchronization, but runs at rates two orders of magnitude faster than the SLM network. The network can be modeled by Eq. \ref{eq:network} with the adjacency matrix shown in Fig. \ref{fig:5node}a.

The cluster state we investigate is depicted in Fig. \ref{fig:5node}a: nodes 0 and 2 are synchronized, nodes 1 and 3 are synchronized, and node 4 is a singleton. The system parameters are $\beta=2.27$, $\sigma=1.17$, and $\delta=0.26$, in agreement with the parameters used to observe the same cluster state in the SLM experiment in ref. \onlinecite{sorrentino2016complete}. An experimental time series of this cluster state is shown in Fig. \ref{fig:5node}b. The background color of each time step indicates the corresponding node (Fig. \ref{fig:5node}a) in the space-time network interpretation. The dotted black lines separate the network time steps $n$. The dynamics of the system in the space-time network interpretation is shown in Fig. \ref{fig:5node}c. In Fig. \ref{fig:5node}b and \ref{fig:5node}c, it is clear that the desired cluster synchronization state is displayed by the experiment.

The network updates at a rate of $F_s/N =$ 2 kHz, where $F_s$= 10 kHz is the system time step and $N=5$ is the number of nodes in the network. This is over two orders of magnitude faster than the $\sim$8 Hz update rate of the SLM experiment of ref. \onlinecite{pecora2014cluster,sorrentino2016complete}.

\subsection{Chimeras in a 5 node, globally-coupled network} 

\begin{figure*}
\centering
\includegraphics[width=\textwidth]{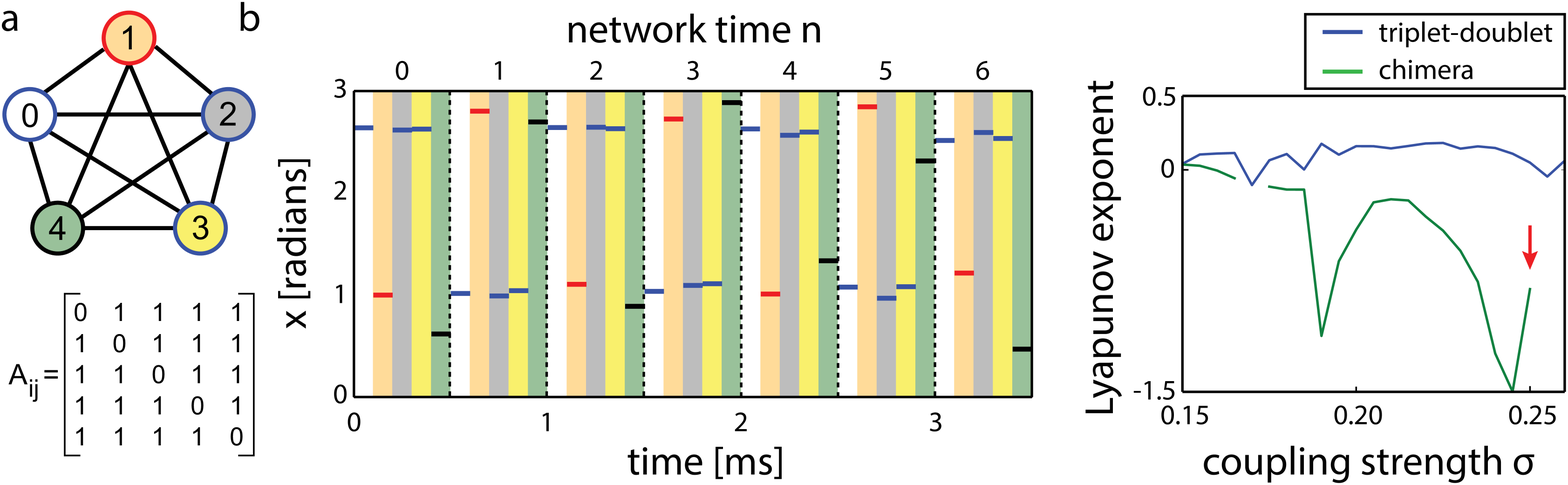}
\caption{\label{fig:5nodechimera}\textit{a}) Depiction of a 5 node globally coupled network. The color of the outline on each node indicates which synchronous cluster it belongs to. \textit{b}) Time series of the 5 node network in the chimera state depicted in \textit{a}. The background color of each time step indicates the node from \textit{a} that is represented by that time step. The coloring of the horizontal bar indicates the synchronous cluster to which that node belongs. The vertical black dotted lines indicate a full network time step $n$. \textit{c}) Stability of chimera state and triplet-doublet cluster state. The red arrow indicates the value of $\sigma$ used to obtain the experimental time series in b. }.
\end{figure*}

The network we discussed in the last section has previously been studied in great detail in ref. \onlinecite{sorrentino2016complete}, and was used to demonstrate the flexibility and speed of our experimental system. We now investigate chimera states in a new network topology.

Experimental observations and stability analysis of chimera states in small networks of four globally-coupled continuous time oscillators have been studied previously \cite{hart2016experimental}. It was predicted that similar chimera states should exist for larger globally coupled networks, and that it should be possible to determine their stability using the analysis techniques described in ref. \onlinecite{pecora2014cluster,hart2016experimental,sorrentino2016complete}. However, due to experimental limitations on the number of nodes and links, chimeras in larger networks of globally coupled oscillators have not been experimentally observed in opto-electronic networks.

We now investigate chimera states in a network of five globally-coupled nodes in our experimental system. The chimera state we consider here is a dynamical state in which three of the nodes are synchronized, and the other two nodes are desynchronized both from the cluster of three and from each other. Group theoretical arguments suggest that such chimera states should exist in globally coupled networks\cite{hart2016experimental}. However, their stability determines whether they should be observable in experiments.

As in ref. \onlinecite{hart2016experimental}, we added a coupling time delay to observe chimera states in our globally-coupled network. This is straightforward to do with the FPGA hardware via shift registers. The network can be modeled as

\begin{equation}
x_i[n+1] = \beta I(x_i[n]) + \sigma\sum_jA_{ij}I(x_j[n-\tau_c]),
\end{equation}
where $I(x)=\sin^2(x+\delta)$ as before and $\tau_c$ is the coupling delay. For this experiment, we take $\tau_c$ = 1 iteration.

Figure \ref{fig:5nodechimera}b shows a typical time series from a chimera in the experiment. The parameter values are $\beta=2.3$, $\sigma=0.25$, and $\delta=\pi/4$. As one can see from Fig. \ref{fig:bif_diagram}, an uncoupled oscillator is chaotic with these parameters. In this chimera, nodes 0, 2, and 3 are in the synchronous (coherent) region, and nodes 1 and 4 are in the desynchronized (incoherent) region. When we start from different initial conditions, different nodes end up in the coherent and incoherent regions. This is expected, since all of our nodes are identical.

In order to investigate the stability of these chimera states, we performed a stability analysis according to the methods described in ref. \onlinecite{pecora2014cluster,hart2016experimental}. We linearize about the chimera state to obtain the variational equations for the network, then use group theory-based techniques to pick out the perturbation directions transverse to the synchronization manifold. The stability of the chimera state is determined by the largest Lyapunov exponent (LLE) of these transverse variational equations: if the LLE is negative, the chimera state is stable. The result of this calculation for $\beta=2.3$ and $\delta=\pi/4$ is shown in Fig. \ref{fig:5nodechimera}c. We see that the chimera state is stable in the region from $\sigma=0.17$ to $\sigma=0.25$. We were not able to calculate the Lyapunov exponent in the transverse direction for some values of $\sigma$. In regions where the chimera state is unstable and a more symmetric state (such as the globally synchronized state or the triplet-doublet state) is stable, one cannot calculate the stability of the chimera state. In this case the trajectory of the chimera state cannot be determined so one cannot linearize about the chimera state and calculate its stability in the usual way.

We also calculate the stability of the triplet-doublet cluster synchronous state, in order to show how the chimera state forms. The results of this calculation are also shown in Fig. \ref{fig:5nodechimera}c. For lower values of $\sigma$, the triplet-doublet cluster state is stable. As $\sigma$ increases, the doublet cluster undergoes isolated desynchronization \cite{pecora2014cluster} and becomes unstable; however, the triplet cluster state remains stable. This results in a triplet-singlet-singlet state, which we call a chimera state.

\section{Conclusions}
We have developed a new experimental approach to study dynamics on large networks with arbitrary topology. This system provides a significant advance in network experiments in terms of speed, cost, size, and flexibility of coupling topologies.
The experiment relies on the space-time interpretation of systems with time delayed feedback in order to create a network of coupled maps out of a single time delayed feedback loop. Because we are able to use a single loop, all of our nodes are truly identical We use optical and electronic components with no highpass filtering in order to remove the nearest-neighbor coupling found in previous systems that rely on the space-time interpretation to create networks\cite{appeltant2011information,larger2012photonic,larger2013virtual,larger2015laser,antonik2017brain}. This, combined with an FPGA in the feedback loop to implement time delays, allows us to study networks with arbitrary topology, with adaptive capabilities, and at high speed. The FPGA also allows us to intentionally introduce heterogeneities in the nodes or links as well as additional sources of noise into our experiments, if desired.

When operated as a single nonlinear map, our experiment has displayed fixed point, periodic, and chaotic behavior for different parameters. We have also experimentally investigated two bidirectionally-coupled maps and networks of coupled maps, both large and small. We have observed cluster synchronization and two different types of chimera states in our system: chimeras in a small globally-coupled network and chimeras in a large ring network. We calculated the stability of the chimera states and cluster states in the 5 node, globally-coupled network, and we showed that the chimera state emerges when the doublet cluster of a triplet-doublet state undergoes isolated desynchronization.

The experiment currently updates at a rate of 10 kHz, which means that a network will update at a rate of 10 kHz/$N$, where $N$ is the number of nodes in the network. Therefore the network updates slowly for extremely large networks (>1000 nodes). Some network experiments with fixed, non-reconfigurable topology operate at much faster time scales; for example, the globally-coupled mode-locked laser system in Ref. \onlinecite{viktorov2014coherence} has dynamics in the range of tens of GHz. However, for all the networks that we consider here, our system provides a significant improvement in speed over previous experiments that allow easily reconfigurable, arbitrary network topology \cite{hagerstrom2012experimental,pecora2014cluster,tinsley2012chimera}. With suitable design it should be possible to increase the clock rate of our experiment many orders of magnitude, allowing experiments on arbitrary networks to occur at unprecedented speeds.

\section*{Acknowledgements} The authors thank Francesco Sorrentino and Lou Pecora for helpful discussions. JDH and RR acknowledge support from ONR Grant No. N000141612481. 

\bibliography{bib}

\end{document}